# Enhanced RMT estimator for signal number estimation in the presence of colored noise

**Huiyue Yi[1, 2], Wuxiong Zhang[1, 2], Hui Xu[1]**

*[1] Science and Technology on Microsystem Laboratory, Shanghai Institute of Microsystem and information technology, Chinese Academy of Science, 4/F, Xinwei Building A, 1455 Pingcheng Road, Jiading, Shanghai, People's Republic of China*

*[2]University of Chinese Academy of Science*

**Email: huiyue.yi@mail.sim.ac.cn; wuxiong.zhang@mail.sim.ac.cn; hui.xu@mail.sim.ac.cn**

## Abstract:

The subspace-based techniques are widely utilized to estimate the parameters of sums of complex sinusoids corrupted by noise, and they need accurate estimation of the signal subspace dimension. The classic RMT estimator for model order estimation based on random matrix theory (RMT) assumes that the noise is white Gaussian, and performs poorly in the presence of colored noise with unknown covariance matrix. In the presence of colored noise, the multivariate regression (MV-R) algorithm models the source detection as a multivariate regression problem and infers the model order from the covariance matrix of the residual error. However, the MV-R algorithm requires that the noise is sufficiently weaker than the signal. In order to deal with these problems, this paper proposes a novel algorithm to estimate the number of signals for the case of colored noise with unknown covariance matrix based on the analysis of the behavior of information theoretic criteria. Firstly, a first criterion is defined as the ratio of the current eigenvalue and the mean of the next ones, and its properties is analyzed with respect to the over-modeling and under-modeling. Moreover, a second criterion is designed as the ratio of the current value and the next value of the first criterion, and its properties is analyzed with respect to the over-modeling and under-modeling. Then, a novel enhanced RMT estimator is proposed for signal number estimation by analyzing the detection properties among the signal number estimates obtained by these two criteria, the MV-R estimator and the RMT estimator to sequentially determine whether the eigenvalue being tested is arising from a signal or from noise. Finally, simulation results are presented to illustrate that the proposed enhanced RMT estimator has better estimation performance and works better than the existing methods both in the case of white noise and in the case of colored noise.

*Index terms*—signal number estimation, model order estimation, colored noise, AIC, random matrix theory

## 1. Introduction

Spectral analysis of signals is a major problem in statistical signal processing, and has wide applications in communications, radar, sonar, seismology, astronomy and so on [1]-[3]. Subspace methods such as MUSIC algorithm [4] and ESPRIT [5] have been applied to estimate the frequencies while overcoming the frequency resolution limitation of the FFT method in low signal-to-noise ratios (SNRs). The subspace methods are based on the eigen-structure of the autocorrelation matrix, and their performances are completely deteriorated by choosing a wrong dimension of the signal subspace or the noise subspace. Therefore, the subspace methods need accurate estimation of the signal subspace dimension, that is, of the number of harmonic components or signals that are superimposed and corrupted by noise. This estimation is particularly difficult when the signal-to-noise ratio (SNR) is low and the statistical properties of the noise are unknown. There are many methods to estimate the number of signals (or the dimension of the signal subspace). The traditional methods are based on the information theoretic criteria (ITC) such as AIC, MDL, etc. [6]-[8]. Utilizing the random matrix theory (RMT), a RMT estimator is proposed to estimate the number of signals via sequentially detecting the largest noise eigenvalues as arising from a signal or noise for a given over-estimation probability [9]. In [10], the linear shrinkage (LS) technique is employed to obtain the estimates for the noise eigenvalues, and the LS-MDL estimator is proposed for signal number estimation. In [11], an improved AIC estimator is proposed for signal number estimation based on the RMT. In [12], a RMT estimator with adaptive decision criterion based on linear shrinkage (LS-RMT-ADC estimator) is proposed for signal number estimation when the number of samples is relatively small. However, these algorithms [6]-[12] assume that the noise is white Gaussian and the signals are uncorrelated, and will perform poorly when the signals are correlated or the noise is not white. In order to cope with this problem, some methods have been developed for determining the number of sources when the noise observations have an unknown covariance matrix [13]-[18]. The signal number estimation method in [14] makes use of two discriminant functions for separating the eigenvalues into two classes, whose separation limit can be found using two discriminant functions. However, this method may be suboptimal because of improper choice of the discriminant functions. The Gerschgörin disk estimator (GDE) criterion in [15] performs better than AIC and MDL for the case of colored noise, but it is less effective for white noise. In the presence of colored noise, the source detection was modelled as a multivariate regression problem and the model order was inferred from the covariance matrix of the residual error [17], and this method will be called as MV-R algorithm for brevity. However, it requires that the noise is sufficiently weaker than the signal. In [16], the extreme value theory (EVT) was utilized to address the problem of model order selection associated with the detection of complex-valued sinusoids in the presence of unknown correlated noise, and a log-likelihood penalty term for EVT-based model selection was derived. In [18], a threshold-based model-order selection method is proposed without assuming precise distribution knowledge of the measurements. Specifically, a Hermitian Toeplitz data matrix is formulated and the signal number is estimated by thresholding the corresponding singular values. Although this method allows an unknown noise covariance matrix, it requires that the noise is sufficiently weaker than the signal. Therefore, its detection performance will degrade when the SNR becomes low.

In order to correctly estimate the dimension of signal subspace for the case of colored noise with unknown covariance matrix, this paper proposes an enhanced RMT estimator to detect the number of sources without assuming precise distribution knowledge about the noise and the signals. The proposed estimator is inspired by the behavior of information theoretic criteria for model order selection in [19]-[20]. Firstly, a first decision criterion is defined as the ratio of the current eigenvalue and the mean of the next ones, and its properties is analyzed with respect to the over-modeling and under-modeling. Secondly, a second decision criterion is designed as the ratio of the current value to the next value of the first decision criterion, and its properties is also analyzed with respect to the over-modeling and under-modeling. Then, an enhanced RMT estimator is proposed for signal number estimation by analyzing the properties among the signal number estimate obtained by the RMT algorithm, the signal number estimate obtained by the MV-R algorithm and indexes of the maximum of the first decision criterion and the second decision criterion to sequentially determine whether the eigenvalue being tested is arising from a signal or from noise. Finally, simulation results are presented to illustrate the estimation performance of the proposed enhanced RMT estimator and compare it with the existing methods.

The remainder of this paper is organized as follows. Section 2 is the problem formulation and the related works that motivate the current work. In Section 3, the novel enhanced RMT estimator for signal number estimation is presented. Simulation results and comparison of the proposed enhanced RMT estimator with other approaches are provided in Section 4, followed by conclusions in Section 5.

## 2. Problem formulation and related works

In this Section, data model and problem formulation are firstly described. Then, the related works that motivate the current work are introduced.

## 2.1. Data model

We consider an array of $M$ antennas and the $n$-th observed $M$-dimensional vector $\mathbf{x}(n)$ [19], which can be expressed as

$$\mathbf{x}(n) = \mathbf{A}\mathbf{s}(n) + \mathbf{w}(n), \quad n = 1, 2, \cdots, N, \tag{1}$$

where $N$ is the number of observations, $\mathbf{A} = [\mathbf{a}_1, \mathbf{a}_2, \cdots, \mathbf{a}_q]$ is the array steering matrix with full column rank, $\mathbf{s}(n) = [s_1(n), s_2(n), \cdots, s_q(n)]^T$ is the vector of $q$ sources modeled as Gaussian random variables. The number of sources $q$ is unknown and has to be estimated. Typically, under the assumption that $\mathbf{w}[n]$ is a zero mean white Gaussian noise with a variance $\sigma^2$, the autocorrelation matrix $\mathbf{R}_{\mathbf{xx}}$ of $\mathbf{x}(n)$ can be written as

$$\mathbf{R}_{\mathbf{xx}} = E[\mathbf{x}(n)\mathbf{x}^H(n)] = \mathbf{A}\mathbf{R}_s\mathbf{A}^H + \sigma^2\mathbf{I}. \tag{2}$$

where $\mathbf{R}_s = E[\mathbf{s}(n)\mathbf{s}^H(n)]$. Assuming that $q$ noise-free population signal eigenvalues of $\mathbf{R}_{\mathbf{xx}}$ is given by $\{\lambda_1, \lambda_2, \cdots, \lambda_q\}$, the population eigenvalues $\{\rho_j\}_{j=1}^{M}$ of $\mathbf{R}_{\mathbf{xx}}$ is given by

$$\{\lambda_1 + \sigma^2, \cdots, \lambda_q + \sigma^2, \sigma^2, \cdots, \sigma^2\}. \tag{3}$$

In practice, the population covariance matrix $\mathbf{R}_{\mathbf{xx}}$ is estimated from a finite number of samples. Given $N$ i.i.d. samples $\{\mathbf{x}(n)\}_{n=1}^{N}$, the sample covariance matrix $\hat{\mathbf{R}}_{\mathbf{xx}}$ can be estimated as

$$\hat{\mathbf{R}}_{\mathbf{xx}} = \frac{1}{N}\sum_{n=1}^{N}\mathbf{x}(n)\mathbf{x}^H(n). \tag{4}$$

Let the sample eigenvalues of $\hat{\mathbf{R}}_{\mathbf{xx}}$ be $l_1 \ge l_2 \ge \cdots \ge l_M$. Then, the problem is to determine the number $q$ of signals from finite noise samples $\{\mathbf{x}(n)\}_{n=1}^{N}$. Assuming that $\mathbf{y}(n)$ in (1) contains only the noise term $\mathbf{w}[n]$, the distribution of the largest eigenvalue of $\hat{\mathbf{R}}_{\mathbf{xx}}$ converges to the Tracy-Widom distribution in the joint limit $p, n \to \infty$ with $p/n \to \gamma \in (0, \infty)$. That is, for every $x \in \mathrm{R}$,

$$\Pr[l_1 < \sigma^2(\mu_{n,p} + x\sigma_{n,p})] \to F_\beta(x). \quad (5)$$

where $\beta = 1$ for real valued noise and $\beta = 2$ for complex-valued noise. Based on this observation, a RMT estimator was proposed in [9] to estimate the number of signals via detecting the largest noise eigenvalues. The RMT estimator is based on a sequence of hypothesis tests, for $k = 1, 2, \cdots, \min(p, n) - 1$,

$$\begin{aligned} &H_k \ : \ \text{at least } k \text{ signals,} \\ &H_{k-1}: \ \text{at most } k-1 \text{ signals.} \end{aligned} \quad (6)$$

For each value of $k$, the noise level $\sigma^2$ and the signal eigenvalue $\{\lambda_i\}_{i=1}^{k}$ are estimated assuming that $l_{k+1}$, $\cdots$, $l_p$ correspond to noise via solutions of the following non-linear system of equations [9]:

$$\sigma^2_{\mathrm{RMT}}(k) - \frac{1}{p-k}\left[\sum_{j=k+1}^{p} l_j + \sum_{j=1}^{k}(l_j - \hat{\rho}_j)\right] = 0, \quad (7)$$

$$\hat{\rho}_j^2 - \hat{\rho}_j\left[l_j + (1 - \frac{p-k}{n})\sigma^2_{\mathrm{RMT}}(k)\right] + l_j\sigma^2_{\mathrm{RMT}}(k) = 0. \quad (8)$$

Then, the likelihood of the $k$th eigenvalue $l_k$ is tested as arising from a signal or from noise as follows:

$$l_k > \sigma^2_{\mathrm{RMT}}(k)\left(\mu_{n,p-k} + s(\alpha)\sigma_{n,p-k}\right). \quad (9)$$

where the value of $s(\alpha)$ depends on the required significance level $\alpha$, and $s(\alpha)$ can be calculated by inverting the Tracy-Widom distribution. If (9) is satisfied, $H_k$ is accepted and $k$ is increased by one. Otherwise, the number of signals is estimated as $\hat{q}_{\mathrm{RMT}} = k - 1$. That is to say

$$\hat{q}_{\mathrm{RMT}} = \arg\min_k \left\{l_k < \sigma^2_{\mathrm{RMT}}(k)\left(\mu_{n,p-k} + s(\alpha)\sigma_{n,p-k}\right)\right\} - 1. \quad (10)$$

However, the RMT estimator performs poorly when the signals are correlated or the noise is not white, and its estimation performance will degrade when the SNR becomes low.

To estimate the number of signals in the presence of colored noise, the source detection was modelled as a multivariate regression problem and a model order estimation algorithm called as the MV-R algorithm was proposed in [17]. The MV-R algorithm was inferred from the covariance matrix of the residual error [17], and its procedure to estimate $q$ is briefly described as follow. Assuming the $N$ sample observations $\mathbf{X} = [\mathbf{x}(1)\,\mathbf{x}(2)\cdots\mathbf{x}(N)]$ is divided into $M_1$ training set $\mathbf{X}_1 = [\mathbf{y}(1)\cdots\mathbf{y}(M_1)]$ and $M_2$ test set $\mathbf{X}_2 = [\mathbf{y}(M_1+1)\cdots\mathbf{y}(N)]$.

For every possible model order $q$ (with $q = 1, 2, \cdots, \min(M, M_1) - 1$), the ML estimates $\hat{\mathbf{B}}_q$ and $\hat{\mathbf{\Sigma}}_q$ are computed by means of Algorithm 1 in [17]. Then, the model order $q$ is estimated as

$$\hat{q}_{\text{MV-R}} = \arg\min_{q}\left[M_2 \log|\hat{\mathbf{\Sigma}}_q| + \mathrm{Tr}(\hat{\mathbf{E}}^H \hat{\mathbf{\Sigma}}_q^{-1}\hat{\mathbf{E}}) + \eta(q)\right], \tag{11}$$

where $\hat{\mathbf{E}} = \mathbf{X}_2 - \mathbf{P}_q\hat{\mathbf{B}}_q$. Please refer to [] for details about the MV-R algorithm. However, the MV-R algorithm requires that the noise is sufficiently weaker than the signal. In order to overcome the problems of the algorithms for the signal number estimation mentioned above, in Section 3 a novel enhanced RMT algorithm for signal number estimation is presented.

## 3. Enhanced RMT algorithm for signal number estimation

Firstly, two criteria are defined based on the analysis of the behavior of information theoretic criteria utilized in model order selection [20]. The first criterion is defined as the ratio of the current eigenvalue and the mean of the next ones, and the second criterion is designed as the ratio of the current value and the next value of the first criterion. Moreover, the properties of these two criteria are analyzed. Then, a novel double-criterion enhanced RMT algorithm is proposed to determine whether the eigenvalue being tested is arising from a signal or from noise by utilizing the relationship among the signal number estimate obtained by the KN algorithm, the signal number estimate obtained by the MV-R algorithm, and the maximum indexes of the first decision criterion and the second decision criterion.

### 3.1. Definition of two criteria and their properties

Firstly, utilizing the eigenvalues $l_1 \geq l_2 \geq \cdots \geq l_M$ of $\hat{\mathbf{R}}_y$ in (4), the arithmetic mean of $l_k, \cdots, l_M$ is defined as

$$A_k = \frac{1}{M-k+1}\sum_{j=k}^{M} l_j \,. \tag{12}$$

As in [20], the fist criterion $d_A(k)$ is defined as the current eigenvalue and the mean of the next ones (i.e., the ratio of $l_k$ to the $A_{k+1}$ as follows:

$$d_A(k) = \frac{l_k}{A_{k+1}}, \qquad k = 1, 2, \cdots, M-1 \,. \tag{13}$$

As is illustrated in [20], the asymptotic estimation bounds of BIC (Bayesian information criterion) and GBIC$_2$ (generalized BIC) are approximately given by the resolvable contrast between the smallest signal sample eigenvalue and the noise variance estimator. According to a well-known theorem [21]-[22], it can be easily inferred that $\sqrt{N} d_A(k) \xrightarrow{d} N(\mu_k, (M-k)^2 b_k^2)$, where $\mu_k$ and $b_k^2$ are respectively given by

$$\mu_k = \sqrt{N}\left[\left(\frac{(M-k)\lambda_k}{\lambda_{k+1}+\cdots+\lambda_M} - 1\right) - O\left((M-k)\sqrt{\frac{c(N)}{N}}\right)\right], \tag{14}$$

$$b_k^2 = \frac{\lambda_k^2}{(\lambda_{k+1}+\cdots+\lambda_M)^2} + \frac{\lambda_k^2\lambda_{k+1}^2}{(\lambda_{k+1}+\cdots+\lambda_M)^4} + \cdots + \frac{\lambda_k^2\lambda_q^2}{(\lambda_{k+1}+\cdots+\lambda_M)^4} + \frac{2\lambda_k^2(M-q)\sigma^4}{(\lambda_{k+1}+\cdots+\lambda_M)^4} \,. \tag{15}$$

In order to ensure correct estimation of the number of signals, the ratio $d_A(q)$ in (13) between the smallest signal sample eigenvalue and the ML estimator of the noise variance should be large, while the ratio $d_A(q+1)$ should be small, where $q$ is the real number of signals. In [20], it is shown that the estimation performance of the AIC and MDL estimators have a close relationship with $d_A(k)$ in (13). With respect to the over-modeling and under-modeling, the following observations are made in [20]: 1) If the signal eigenvalues $l_1, \cdots, l_q$ are of about the same order of magnitude and are well separated from the noise eigenvalues $l_{q+1}, \cdots, l_M$, then under-modeling is very unlikely to happen; 2) If the noise eigenvalues are not clustered sufficiently closely, then the ITC may ignore an arbitrarily large gap between the signal and the noise eigenvalues, leading to over-modeling; 3) If the signal and the noise eigenvalues are not well separated and if the noise eigenvalues are clustered sufficiently closely, then under-modeling is likely to happen. Due to these reasons, $d_A(k)$ in (13) cannot be utilized as a criterion for signal number estimation, and the properties of the $d_A(k)$ should be further analyzed.

As to the index $k$ of the maximum value of $d_A(k)$, we obtain the following novel observations:

(a) When there are no signals (i.e., $q=0$), the index $k$ of the maximum value of $d_A(k)$ may be a random number in the range $[1, M-1]$.

(b) When there are signals (i.e., $q \geq 1$**), t**he values of $d_A(k)$ are all relatively large and of the same order of magnitude for $k \leq q$, while the values of $d_A(k)$ are all relatively small and of the same order of magnitude for $k \geq q+1$ when the SNR is relatively small. However, when the SNR becomes larger, the values of $d_A(k)$ are relatively large and of the same order of magnitude for $k \leq q+1$, while the values of $d_A(k)$ are relatively small and of the same order of magnitude for $k \geq q+2$. In this case, the noise eigenvalue may wrongly detected as the signal eigenvalue if the index $k$ of the maximum value of $d_A(k)$ is utilized as the decision criterion to determine whether $l_k$ is arising from signal.

Based on above observations (a) and (b) and with the aim to make a clear discrimination between the smallest signal eigenvalue and the largest noise eigenvalue, the second criterion is proposed and defined as follows:

$$d_{A,A}(k) = \frac{d_A(k)}{d_A(k+1)}, \qquad k = 1, 2, \cdots, M-2. \tag{16}$$

As the index $k$ of the maximum value of $d_{A,A}(k)$ versus SNR is analyzed, and the following observations are obtained:

(c) When there are no signals (i.e., $q=0$), the index $k$ of the maximum value of $d_{A,A}(k)$ is a random number in the range $[1, M-2]$.

(d) When there are signals (i.e., $q \ge 1$), the $d_{A,A}(k)$ will be maximized at $k = q$ when the SNR is relatively small, and will be maximized at $k = q+1$ when the SNR becomes large.

**3.2. Proposed enhanced RMT estimator for signal number estimation**

Utilizing the properties of the signal number estimate $\hat{q}_{\text{RMT}}$ in (10) obtained by the RMT algorithm, the signal number estimate $\hat{q}_{\text{MV-R}}$ obtained by the MV-R algorithm, the maximum index of the first decision criterion $d_A(k)$ in (13) and the maximum index of the second criterion $d_{A,A}(k)$ in (16), a novel enhanced RMT estimator is proposed to estimate the signal number $\hat{q}_{\text{est}}$ according to the following steps:

**Step 1:** Estimate the number of signals $\hat{q}_{\text{RMT}}$ using (10).

**Step 2:** Estimate the number of signals $\hat{q}_{\text{MV-R}}$ using (11).

**Step 3:** Estimate the index of the maximum value of $d_A(k)$ in (13) as:

$$\hat{k}_1 = \arg \max_{k=1,\cdots,M-1} d_A(k). \tag{17}$$

**Step 4**: Estimate the index of the maximum value of $d_{A,A}(k)$ in (16) as:

$$\hat{k}_2 = \arg \max_{k=1,\cdots,M-2} d_{A,A}(k). \tag{18}$$

**Step 5:** Using the relationship among $\hat{q}_{\text{RMT}}$, $\hat{q}_{\text{MV-R}}$, $\hat{k}_1$ and $\hat{k}_2$, the number of signals is estimated in the following way:

(1) If $\hat{k}_2 = \hat{q}_{\text{MV-R}}$ and $\hat{k}_2 = \hat{q}_{\text{RMT}}$, it can be inferred that the eigenvalue $l_1$ is arising from a signal. Therefore, the number of signals is estimated as $\hat{q} = \hat{k}_2$. Then, go to Step 6.

(2) Otherwise, if $\hat{k}_2 = \hat{q}_{\text{RMT}}$, it can be inferred that $l_1$ is arising from a signal. Therefore, the number of signals is estimated as $\hat{q} = \hat{q}_{\text{RMT}}$. Then, go to Step 6.

(3) Otherwise, if $\hat{q}_{\text{RMT}} = \hat{q}_{\text{MV-R}}$, the number of signals is estimated as follows:

1) If $\hat{k}_2 = \hat{k}_1$ and $\hat{k}_2 = \hat{q}_{\mathrm{RMT}} - 1$, the number of signals is estimated as $\hat{q} = \hat{q}_{\mathrm{RMT}} - 1$. Then, go to Step 6.

2) Otherwise, the number of signals is estimated as $\hat{q} = \hat{q}_{\mathrm{RMT}}$. Then, go to Step 6.

(4) Otherwise, if $\hat{k}_2 = \hat{q}_{\text{MV-R}}$, the number of signals is estimated in the following way:

1) if $\hat{q}_{\mathrm{RMT}} = 0$, the number of signals is estimated as $\hat{q} = \hat{q}_{\mathrm{RMT}}$. Then, go to Step 6.

2) the number of signals is estimated as $\hat{q} = \hat{k}_2$. Then, go to Step 6.

(5) Otherwise, the number of signals is estimated in the following way:

1) If $\hat{k}_2 = \hat{k}_1$, the number of signals is estimated as follows:

(a) If $\hat{q}_{\mathrm{RMT}} = 0$ and $\hat{q}_{\text{MV-R}} = 1$, the number of signals is estimated as $\hat{q} = \hat{q}_{\mathrm{RMT}}$. Then, go to Step 6.

(b) Otherwise, the number of signals is estimated as $\hat{q} = \hat{k}_2$. Then, go to Step 6.

2) Otherwise, the number of signals is estimated in the following way:

(a) If $|\hat{k}_2 - \hat{k}_1| \leq 1$, the number of signals is estimated in the following way:

a) If $\hat{k}_1 = \hat{k}_2$, it can be inferred that the eigenvalue $l_{\hat{k}_1}$ is arising from a signal. Therefore, the number of signals is estimated as $\hat{q} = \hat{k}_2$. Then, go to Step 6.

b) Otherwise, i.e., $\hat{k}_2 \neq \hat{k}_1$, the number of signals is estimated in the following way:

① If $\hat{q}_{\mathrm{RMT}} > \hat{q}_{\text{MV-R}}$, it can be inferred that $\hat{q}_{\mathrm{RMT}}$ may over-estimate the number of signals. Therefore, the number of signals is estimated as $\hat{q} = \hat{q}_{\text{MV-R}}$. Then, go to Step 6.

② Otherwise, i.e., $\hat{q}_{\mathrm{RMT}} \leq \hat{q}_{\text{MV-R}}$, it can be inferred that $\hat{q}_{\text{MV-R}}$ may over-estimate the number of signals. Therefore, the number of signals is estimated as $\hat{q} = \hat{q}_{\mathrm{RMT}}$. Then, go to Step 6.

(b) Otherwise, i.e., $|\hat{k}_2 - \hat{k}_1| > 1$, the number of signals is estimated in the following way:

a) If $|\hat{q}_{\mathrm{RMT}} - \hat{q}_{\text{MV-R}}| \leq 1$, the number of signals is estimated in the following way:

① If $\hat{q}_{\mathrm{RMT}} > \hat{q}_{\text{MV-R}}$, the number of signals is estimated in the following way:

I) If $\hat{q}_{\mathrm{RMT}} > \hat{k}_2$, the number of signals is estimated as $\hat{q} = \hat{q}_{\mathrm{RMT}}$. Then, go to Step 6.

II) Otherwise, i.e. $\hat{q}_{\mathrm{RMT}} \leq \hat{k}_2$, the number of signals is estimated as $\hat{q} = \hat{k}_2$. Then, go to Step 6.

② Otherwise, i.e., $\hat{q}_{\text{RMT}} \leq \hat{q}_{\text{MV-R}}$, the number of signals is estimated in the following way:

I) If $\hat{q}_{\text{RMT}} = 0$ and $\hat{q}_{\text{MV-R}} = 1$, the number of signals is estimated as $\hat{q} = 0$. Then, go to Step 6.

II) Otherwise, the number of signals is estimated in the following way:

i) If $\hat{q}_{\text{MV-R}} \geq \hat{k}_2$, the number of signals is estimated as $\hat{q} = \hat{q}_{\text{MV-R}}$. Then, go to Step 6.

ii) Otherwise, i.e., $\hat{q}_{\text{MV-R}} < \hat{k}_2$, number of signals is estimated as $\hat{q} = \hat{k}_2$. Then, go to Step 6.

b) Otherwise, i.e., $|\hat{q}_{\text{RMT}} - \hat{q}_{\text{MV-R}}| > 1$, the number of signals is estimated in the following way:

① If $\hat{q}_{\text{RMT}} > \hat{q}_{\text{MV-R}}$, the number of signals is estimated as $\hat{q} = \hat{q}_{\text{RMT}}$. Then, go to Step 6.

② Otherwise, i.e., $\hat{q}_{\text{RMT}} \leq \hat{q}_{\text{MV-R}}$, the number of signals is estimated as $\hat{q} = \hat{q}_{\text{MV-R}}$. Then, go to Step 6.

**Step 6:**

Obtain the signal number estimate $q_{\text{est}} = \hat{q}$.

## 4. Simulation results

In this section, simulations are carried out to evaluate the detection performance of the proposed double-criterion enhanced RMT estimator for signal number estimation and compare it with the existing methods including the conventional AIC estimator [6], the conventional MDL estimator [7], the RMT estimator in [9]. the LS-MDL estimator in [10], the MAIC estimator in [11], and the MV-R algorithm in [17]. As in [17], the matrix $\mathbf{A}$ in the data model in (1) is assumed to be unitary, and $q$ random values uniformly distributed between 1 and 10 are generated as the signal eigenvalues $\{\lambda_i\}_{i=1}^{q}$. The noise is simulated in the same way as in [17], randomly choose $\mathbf{R}_{nn} = M \dfrac{\mathbf{U\Omega U}^H}{\text{Tr}\{\mathbf{\Omega}\}}$. The trace of $\mathbf{R}_{nn}$ is $M$, $\mathbf{U}$ is uniformly distributed on the unitary group, and $\mathbf{\Omega}$ is a diagonal matrix of $M$ eigenvalues uniformly distributed between 0.1 and 1. Different structures of covariance matrices $\mathbf{\Sigma} = E[\mathbf{w}(t)\mathbf{w}^H(t)]$ of are considered to address three models of the noise $\mathbf{w}[n]$: white noise $\mathbf{\Sigma}_{w1} = \mathbf{I}$; a mixture of white and colored noise $\mathbf{\Sigma}_{w2} = \mathbf{I} + \mathbf{R}_{nn}$; colored noise $\mathbf{\Sigma}_{w3} = \mathbf{R}_{nn}$. the signal-to-noise ratio (SNR) is defined as the ratio of the smallest signal eigenvalue in $\{\lambda_i\}_{i=1}^{q}$ and the largest noise eigenvalue in $\mathbf{\Sigma}$. In addition, the performance measure for the signal number estimation is the correct estimation probability defined as

$$P_{CE} = \Pr(q_{\text{est}} = q). \tag{16}$$

The parameters utilized in the simulations are set as. The number of antennas is set as $M = 50$, the samples is set as $N = 150$, and the number of signals is set as $q = 8$. The block size for the MV-R algorithm in [17] is set as $s = 2$, the significance level $\alpha$ of the RMT estimator in [9] is set as $\alpha = 0.005$. In all simulations, the results of various methods are based on 1000 independent trials.

### 4.1. Performance of the proposed enhanced RMT estimator for the case of $q = 0$

In this simulation, the case for $q = 0$ is considered, i.e., there are no signals. In this simulation, the results of various methods are based on 1000 independent trials. Table 1 shows the correct estimation probability $P_{CE}$ of various algorithms for three models of the noise $\mathbf{w}[n]$. As can be seen from Table 1, the MV-R estimator in [17] fails to work properly for three models of the noise $\mathbf{w}[n]$ when $q = 0$. The RMT estimator works properly for the noise models $\mathbf{\Sigma}_{w1} = \mathbf{I}$ and $\mathbf{\Sigma}_{w2} = \mathbf{I} + \mathbf{R}_{nn}$. Moreover, the proposed enhanced RMT estimator works properly for the noise models $\mathbf{\Sigma}_{w1} = \mathbf{I}$ and $\mathbf{\Sigma}_{w2} = \mathbf{I} + \mathbf{R}_{nn}$, and exhibits the same estimation performance as the RMT estimator. Therefore, the proposed enhanced RMT estimator outperforms the MV-R estimator, and has the same estimation performance as the RMT estimator.

**Table 1.** Correct estimation probability $P_{CE}$ of various algorithms for three models of the noise $\mathbf{w}[n]$

| Models of the noise | $\mathbf{\Sigma}_{w1} = \mathbf{I}$ | $\mathbf{\Sigma}_{w2} = \mathbf{I} + \mathbf{R}_{nn}$ |
|---|---|---|
| Enhanced RMT | 1.0000 | 0.9630 |
| AIC | 0.9995 | 0.9550 |
| MDL | 1.0000 | 1.0000 |
| LS-MDL | 0.3800 | 0.0020 |
| MAIC | 1.0000 | 1.0000 |
| RMT | 1.0000 | 0.9670 |
| MV-R | 0.0000 | 0.0000 |

**4.2. Performance of the proposed enhanced RMT estimator for the case of $q=1$**

In this simulation, the case for $q=1$ is considered, i.e., there is one signal. In this simulation, the results of various methods are based on 1000 independent trials. Figure 1 shows the probability of correct estimation of various algorithms versus SNR for the white noise $\mathbf{\Sigma}_{w1}=\mathbf{I}$ when $q=1$, Figure 2 shows the corresponding results for the white and colored noise model $\mathbf{\Sigma}_{w2}=\mathbf{I}+\mathbf{R}_{nn}$, and Figure 3 shows the corresponding results for the colored noise model $\mathbf{\Sigma}_{w3}=\mathbf{R}_{nn}$. As can be seen from Figure 1 for the white noise $\mathbf{\Sigma}_{w1}=\mathbf{I}$, the proposed enhanced RMT estimator has nearly the same estimation performance as the RMT estimator and the AIC estimator and inferior estimation performance to the LS-MDL and MV-R algorithm, while has better estimation performance than the MDL algorithm and the MAIC algorithm. As can be seen from Figure 2 for the mixture of white and colored noise $\mathbf{\Sigma}_{w2}=\mathbf{I}+\mathbf{R}_{nn}$, the proposed enhanced RMT estimator is inferior to the MV-R algorithm. Although the AIC and the LS-MDL estimator outperform the proposed enhanced RMT estimator when the SNR is relatively low, but they have non-neglectable over-estimation probability when the SNR becomes larger. Moreover, the proposed enhanced RMT estimator has the same estimation performance as the RMT estimator, and has better estimation performance than the MDL and MAIC algorithms in this case. As can be seen from Figure 3 for the colored noise $\mathbf{\Sigma}_{w3}=\mathbf{R}_{nn}$, the proposed enhanced RMT estimator outperforms all other algorithms including the AIC, MDL, LS-MDL, MAIC, RMT estimator and the MV-R estimator.

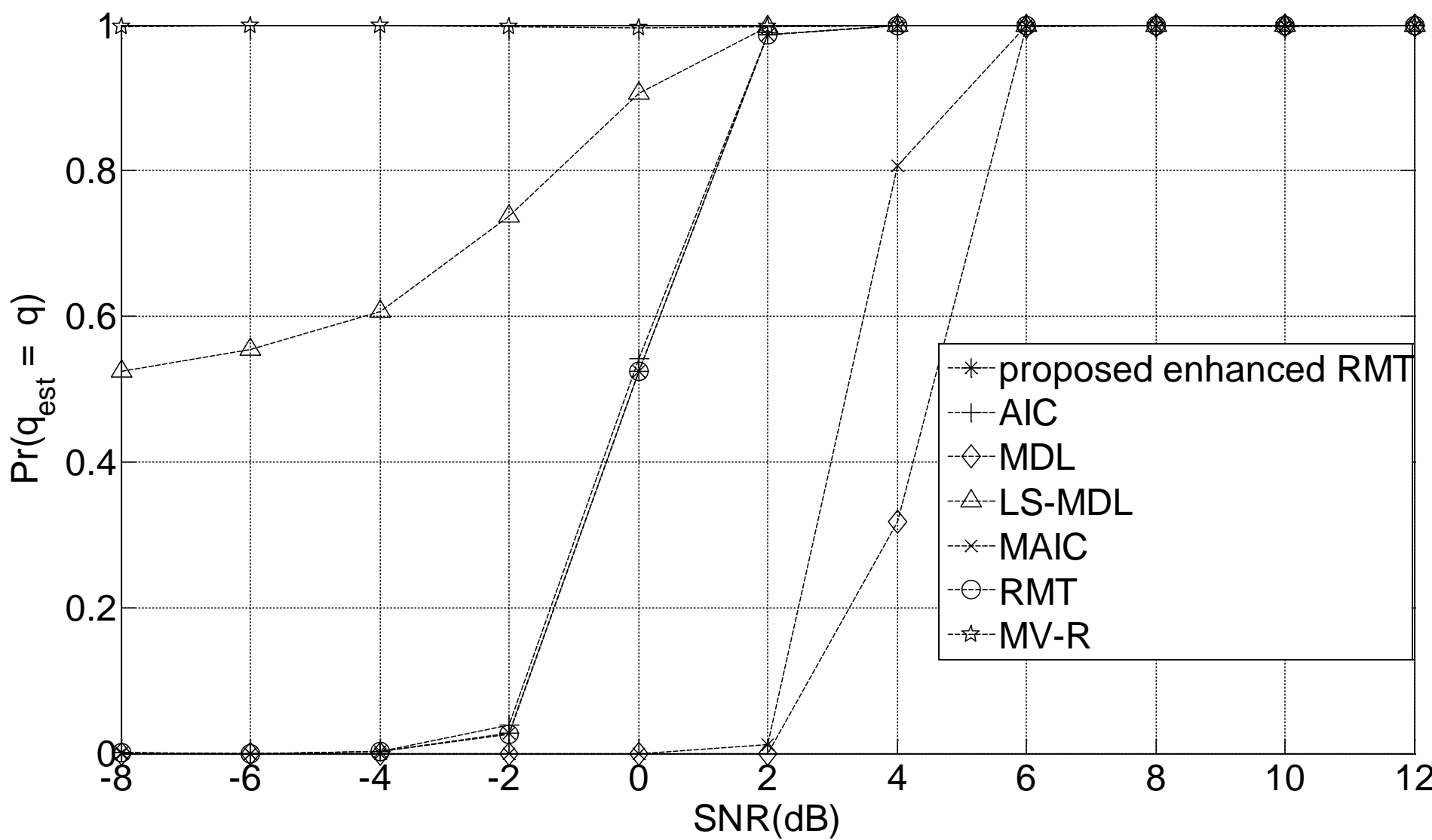

Fig. 1 Probability of correct estimation of various algorithms versus SNR for the white noise $\mathbf{\Sigma}_{w1} = \mathbf{I}$ when $q = 1$.

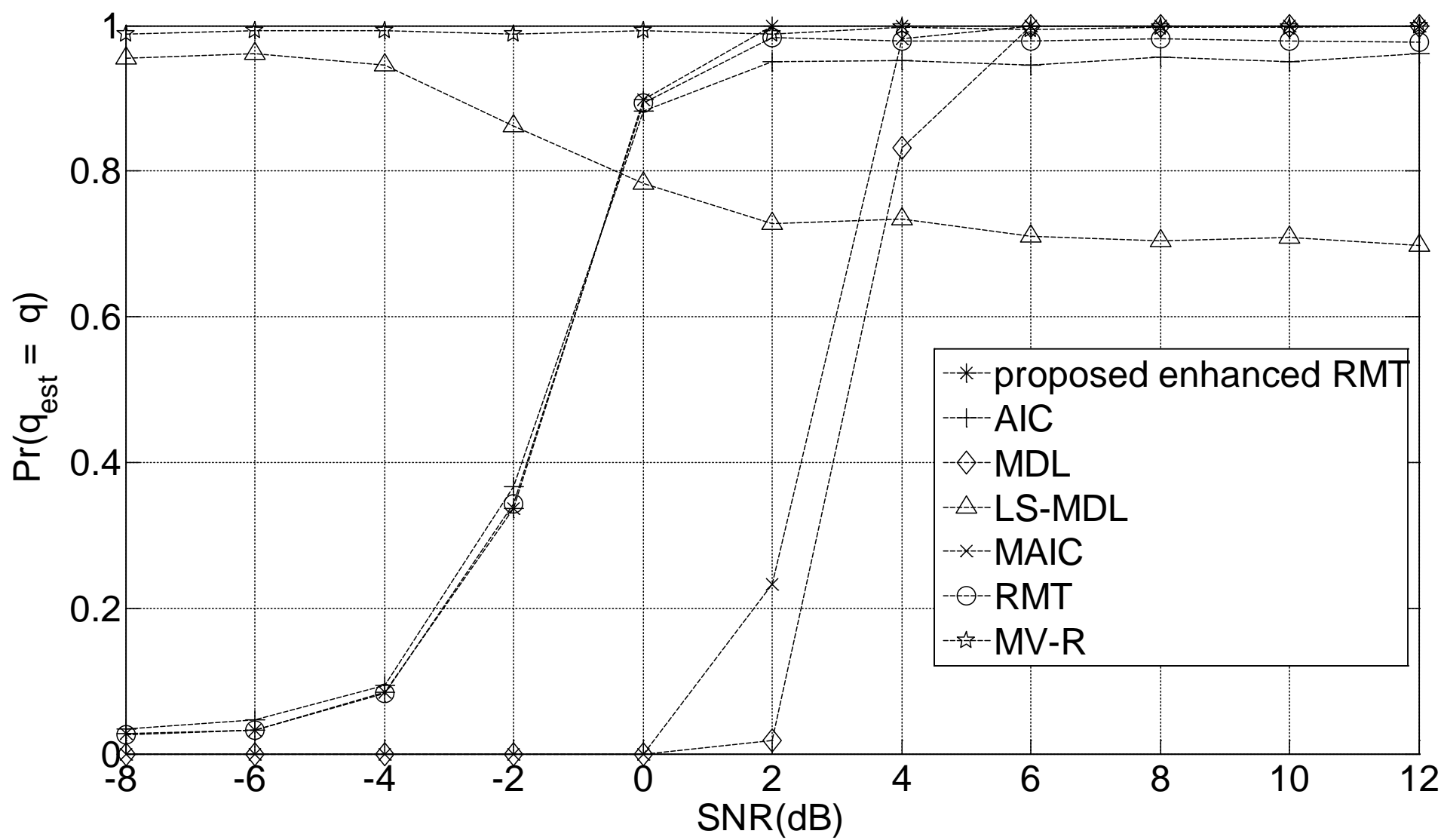

Fig. 2 Probability of correct estimation of various algorithms versus SNR for the mixture of white and colored noise $\mathbf{\Sigma}_{w2} = \mathbf{I} + \mathbf{R}_{nn}$ when $q = 1$.

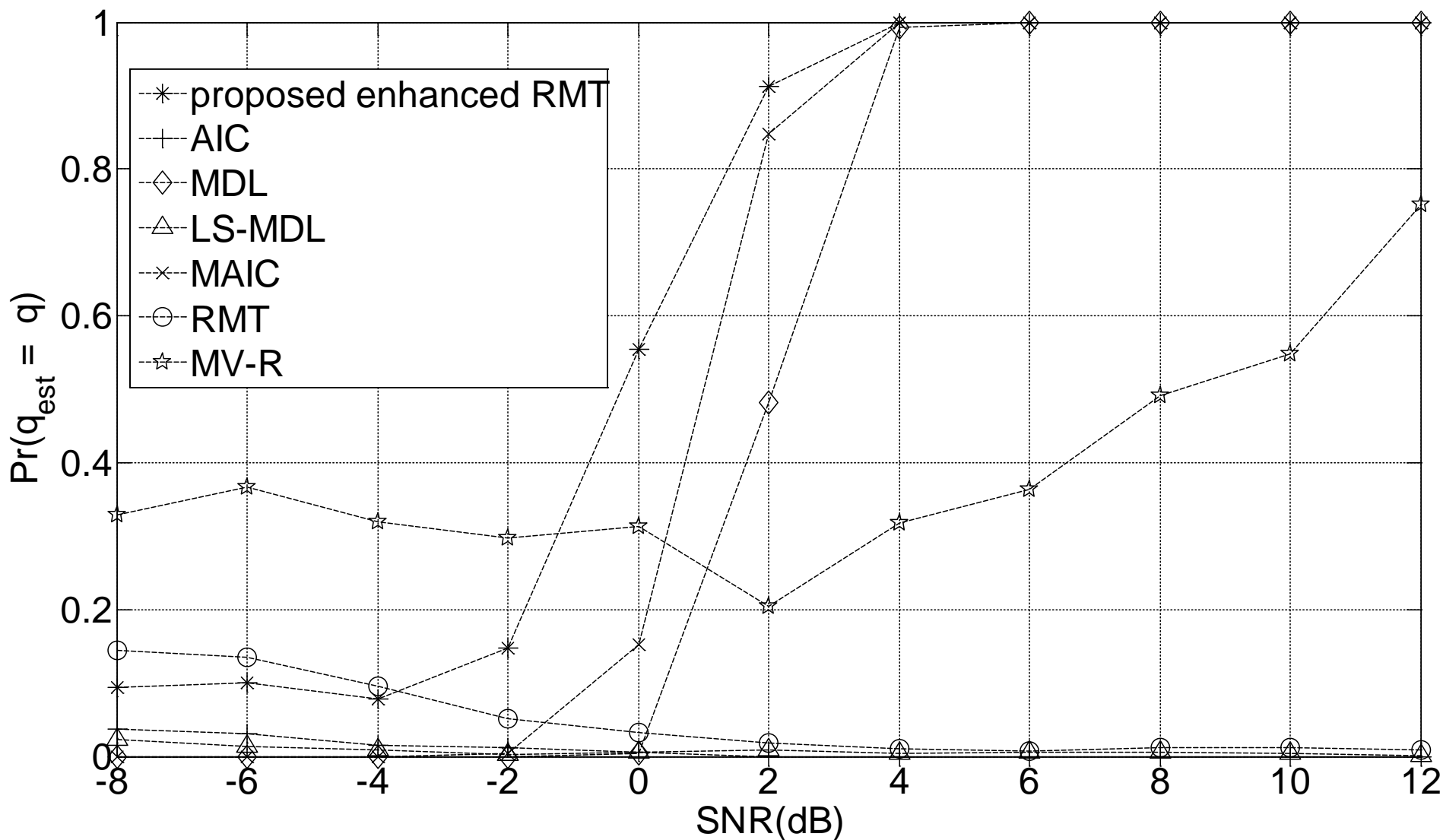

Fig. 3 Probability of correct estimation of various algorithms versus SNR for the colored noise $\mathbf{\Sigma}_{w3} = \mathbf{R}_{nn}$ when $q = 1$.

### 4.3. Performance of the proposed enhanced RMT estimator for the case of $q = 10$

In this simulation, the case for $q = 10$ is considered, i.e., there is eight signals. In this simulation, the results of various methods are based on 1000 independent trials. Figure 4 shows the probability of correct estimation of various algorithms versus SNR for the white noise model $\mathbf{\Sigma}_{w1} = \mathbf{I}$ when $q = 10$, Figure 5 shows the corresponding results for the mixture of white and colored noise $\mathbf{\Sigma}_{w2} = \mathbf{I} + \mathbf{R}_{nn}$, and Figure 6 shows the corresponding results for the colored noise $\mathbf{\Sigma}_{w3} = \mathbf{R}_{nn}$. As can be seen from Figure 4 and Figure 5 for the noise models $\mathbf{\Sigma}_{w1} = \mathbf{I}$ and $\mathbf{\Sigma}_{w2} = \mathbf{I} + \mathbf{R}_{nn}$, the correct estimation probability of the proposed enhanced RMT estimator is slightly less than that of the RMT estimator, but is far higher than that of the other algorithms including the MDL, the LS-MDL, the MAIC, the RMT, and the MV-R estimator. In addition, though the AIC estimator has higher correct estimation probability than the proposed enhanced RMT estimator when the SNR is relatively small, but it has non-neglectable over-estimation probability when the SNR becomes larger. As can be seen from Figure 6 for the noise model $\mathbf{\Sigma}_{w3} = \mathbf{R}_{nn}$, the correct estimation probability of the proposed enhanced RMT estimator is slightly higher than that of the MV-R estimator, while the RMT estimator fails to work properly in this case. Although the LS-MDL estimator has higher correct estimation probability than the proposed enhanced RMT estimator, it has non-neglectable over-estimation probability when the SNR becomes larger. In summary, the proposed enhanced RMT estimator has higher correct estimation probability than the MV-R estimator for all three noise models $\mathbf{\Sigma}_{w1} = \mathbf{I}$ , $\mathbf{\Sigma}_{w2} = \mathbf{I} + \mathbf{R}_{nn}$ and $\mathbf{\Sigma}_{w3} = \mathbf{R}_{nn}$ . Moreover, the proposed enhanced RMT estimator has nearly the same correct estimation probability as the proposed enhanced RMT estimator for the noise models $\mathbf{\Sigma}_{w1} = \mathbf{I}$ and $\mathbf{\Sigma}_{w2} = \mathbf{I} + \mathbf{R}_{nn}$ , while it outperforms the RMT estimator for the noise model $\mathbf{\Sigma}_{w3} = \mathbf{R}_{nn}$ .

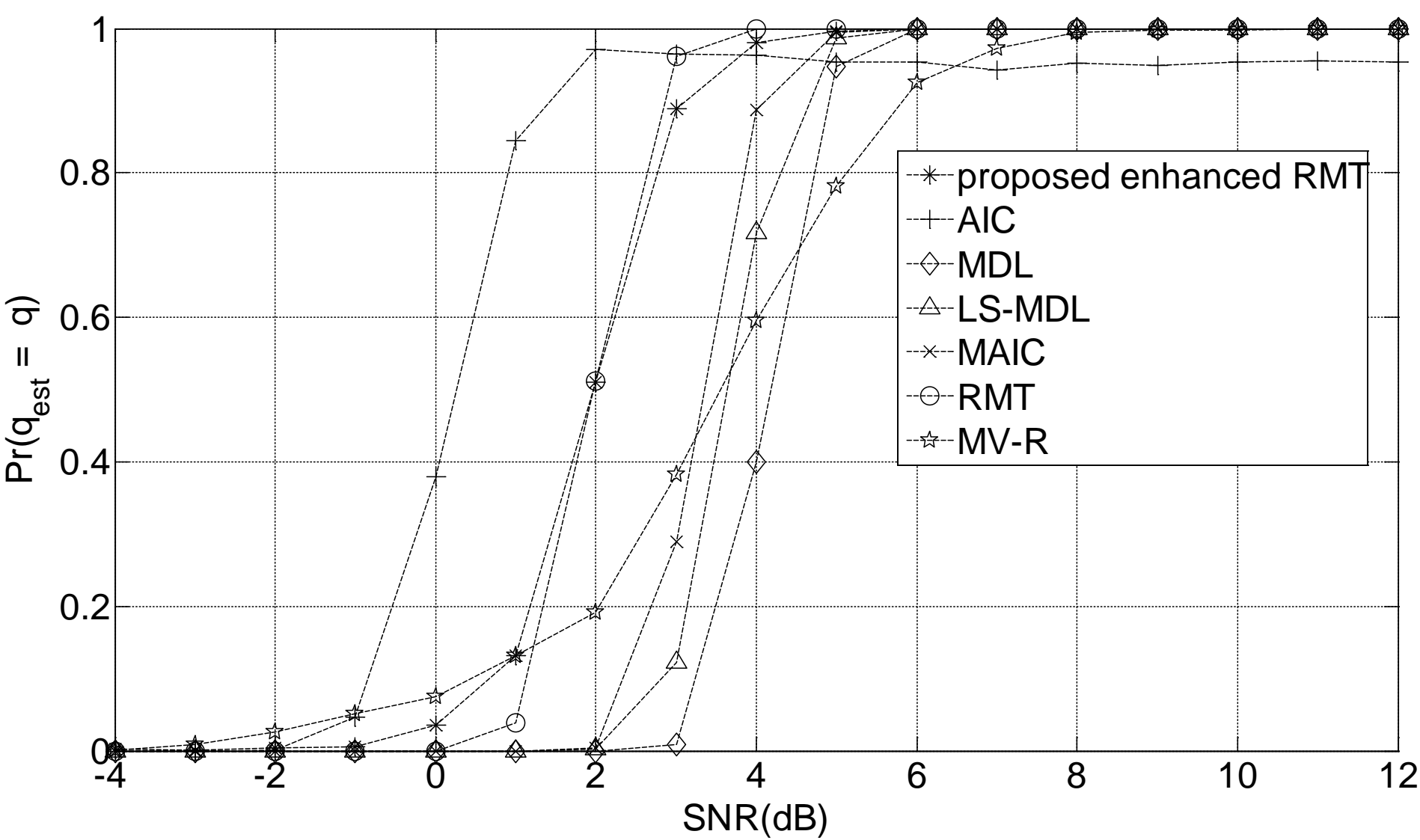

Fig. 4 Probability of correct estimation of various algorithms versus SNR for the white noise $\mathbf{\Sigma}_{w1} = \mathbf{I}$ when $q = 10$ .

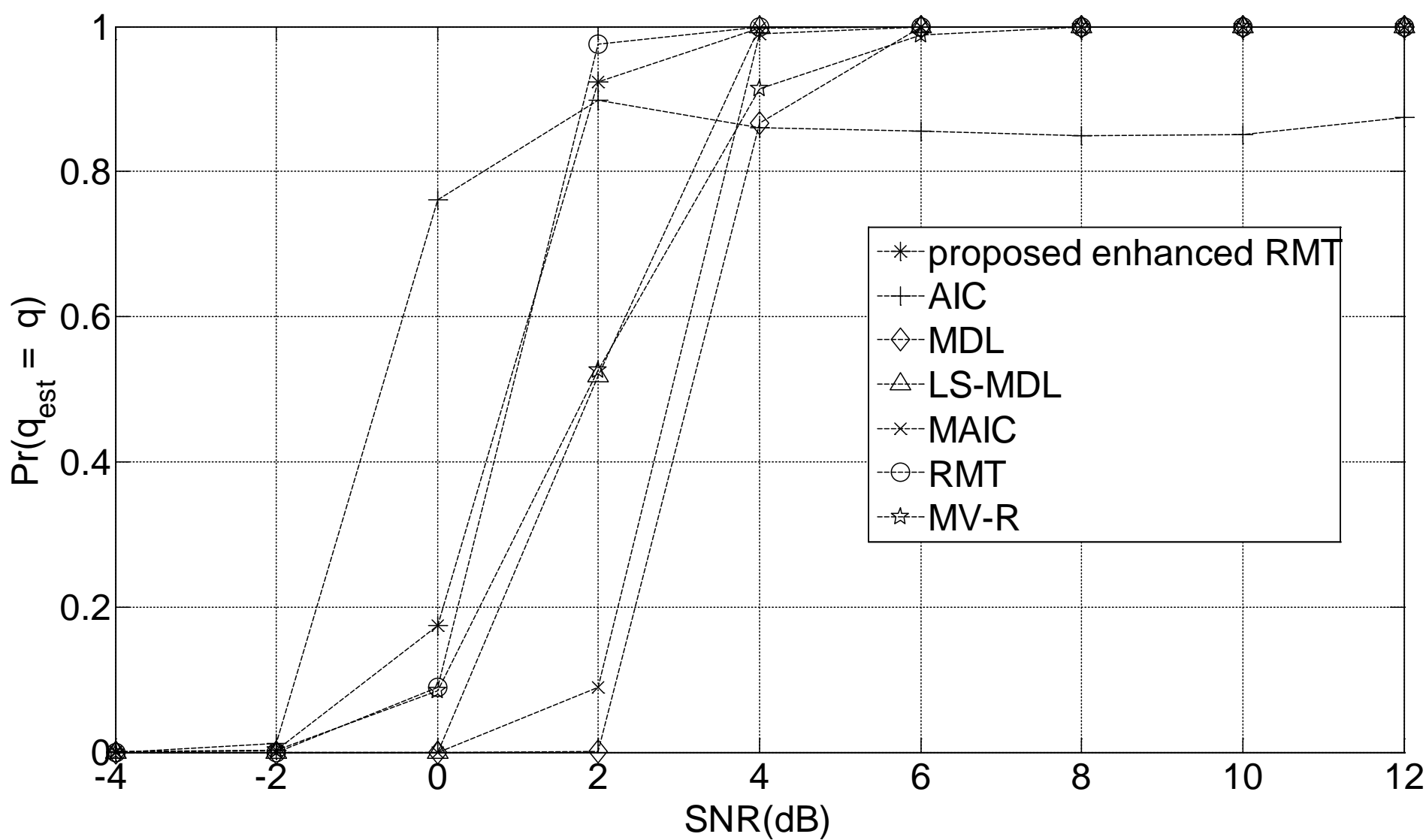

Fig. 5 Probability of correct estimation of various algorithms versus SNR for the mixture of white and colored noise $\mathbf{\Sigma}_{w2} = \mathbf{I} + \mathbf{R}_{nn}$ when $q = 10$.

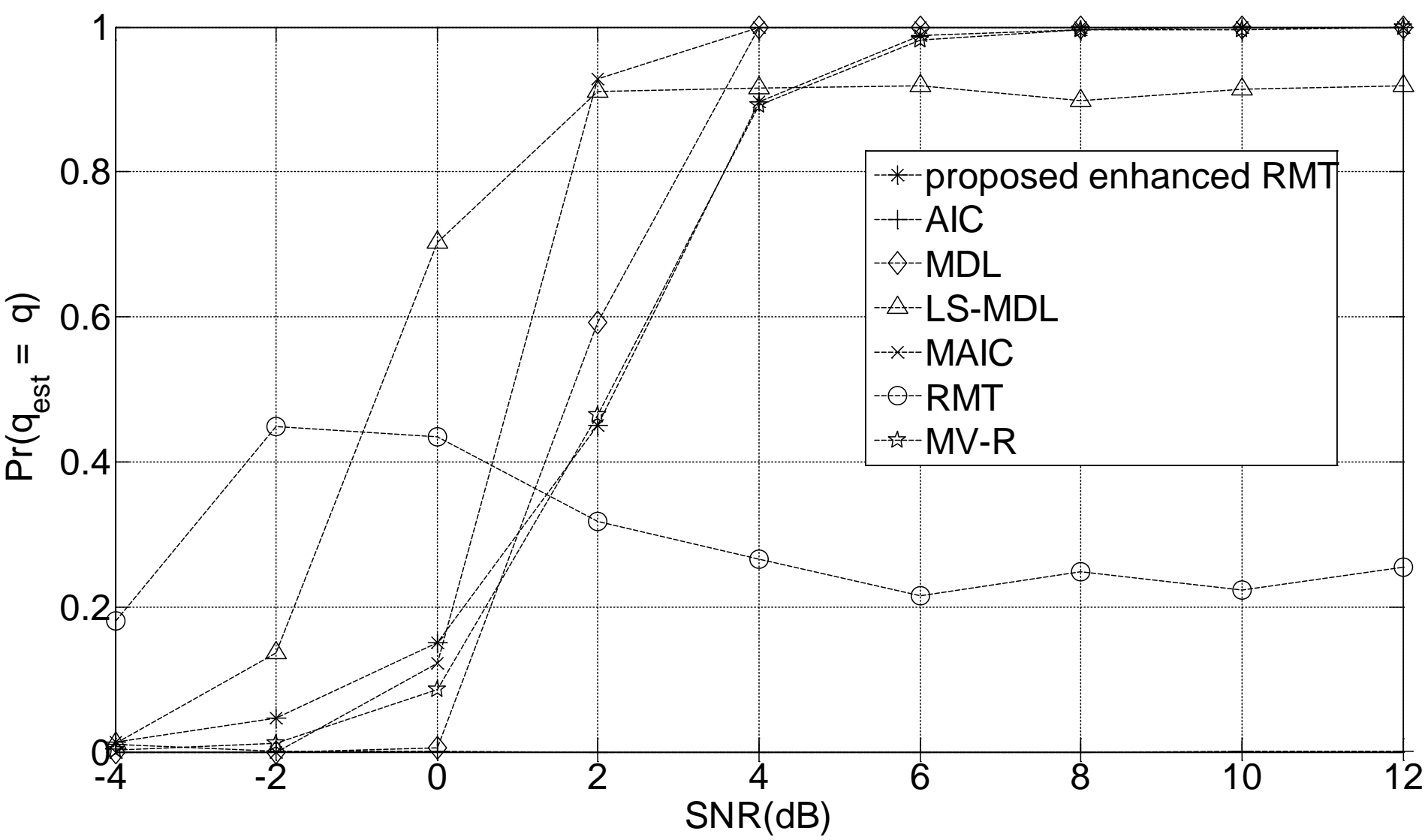

Fig. 6 Probability of correct estimation of various algorithms versus SNR for the colored noise $\mathbf{\Sigma}_{w3} = \mathbf{R}_{nn}$ when $q = 10$.

## 4. Conclusions

In this paper, an enhanced RMT estimator has been proposed to estimate the number of signals for the case of colored noise with unknown covariance matrix. The proposed strategy is based on the analysis of the behavior of information theoretic criteria utilized in the model order selection. Firstly, a first criterion is defined as the ratio of the current eigenvalue and the mean of the next ones, and its properties is analyzed with respect to the over-modeling and under-modeling. Secondly, a novel second criterion is designed as the ratio of the current value and the next value of the first criterion, and its properties is also analyzed with respect to the over-modeling and under-modeling. Then, a novel enhanced RMT estimator is proposed for signal number estimation by utilizing these two criteria, the MV-R estimator and the RMT estimator to sequentially determine whether the eigenvalue being tested is arising from a signal or from noise. Finally, simulation results have been presented to show that the proposed enhanced RMT estimator has better estimation performance than the existing signal number estimation methods, and achieves an optimal tradeoff between the RMT estimator and the MV-R estimator. Moreover, the proposed enhanced RMT estimator works well both in the case of white noise and in the case of colored noise.

## 5. Acknowledgement

This work was supported by the Key Research and Development Task Special Project of the Xinjiang Uygur Autonomous Region (grant number 2022B01009).